\documentclass[useAMS,usenatbib]{mn2e}

\usepackage{verbatim,graphicx,epsfig,dcolumn}
\newcolumntype{.}{D{.}{.}{4}}
\newcolumntype{,}{D{.}{.}{2}}
\newcolumntype{;}{D{.}{.}{1}}
\newcommand{\nodata}{$\cdot\cdot\cdot$}
\newcommand{\lesssim}{{\lower-1.2pt\vbox{\hbox{\rlap{$<$}\lower5pt\vbox{\hbox{$\sim$}}}}}}
\newcommand{\gtrsim}{{\lower-1.2pt\vbox{\hbox{\rlap{$>$}\lower5pt\vbox{\hbox{$\sim$}}}}}}

\title[Dust production in $\omega$ Centauri]{\protect{\emph{Spitzer}} spectra of evolved stars in $\omega$ Centauri and their low-metallicity dust production}
\author[I. McDonald et al.]{I.~McDonald$^{1,2}$\thanks{E-mail:
mcdonald@jb.man.ac.uk}, J.~Th.~van Loon$^{2}$, G.~C.~Sloan$^{3}$, A.~K.~Dupree$^{4}$, A.~A.~Zijlstra$^{1}$, \newauthor
M.~L.~Boyer$^{5}$, R.~D.~Gehrz$^{6}$, A.~Evans$^{2}$, C.~E.~Woodward$^{6}$, C.~I.~Johnson$^{7,8}$\\
$^{1}$Jodrell Bank Centre for Astrophysics, Alan Turing Building, Manchester, M13 9PL, UK\\
$^{2}$Lennard-Jones Laboratories, Keele University, Staffordshire, ST5 5BG, UK\\
$^{3}$Cornell University, Astronomy Department, Ithaca, NY 14853-6801, USA\\
$^{4}$Harvard-Smithsonian Center for Astrophysics, 60 Garden Street, Cambridge, MA 02138, USA\\
$^{5}$STScI, 3700 San Martin Drive, Baltimore, MD 21218, USA\\
$^{6}$Department of Astronomy, 116 Church Street SE, University of Minnesota, Minneapolis, MN 55455, USA\\
$^{7}$Department of Physics and Astronomy, UCLA, 430 Portola Plaza, Box 951547, Los Angeles, CA 90095-1547, USA\\
$^{8}$National Science Foundation Astronomy and Astrophysics Postdoctoral Fellow}

\begin{document}

\date{Accepted 9999 December 32. Received 9999 December 32; in original form 9999 December 32}

\pagerange{\pageref{firstpage}--\pageref{lastpage}} \pubyear{9999}

\maketitle

\label{firstpage}

\begin{abstract}
Dust production is explored around 14 metal-poor ([Fe/H] = --1.91 to --0.98) giant stars in the Galactic globular cluster $\omega$ Centauri using new \emph{Spitzer} IRS spectra. This sample includes the cluster's post-AGB and carbon stars and is thus the first representative spectral study of dust production in a metal-poor ([Fe/H] $< -1$) population. Only the more metal rich stars V6 and V17 ([Fe/H] = --1.08, --1.06) exhibit silicate emission, while the five other stars with mid-infrared excess show only a featureless continuum which we argue is caused by metallic iron dust grains. We examine the metallicity of V42, and find it is likely part of the metal-rich population ([Fe/H] $\sim$ --0.8). Aside from the post-AGB star V1, we find no star from the cluster's bulk, metal-poor ([Fe/H] $< -1.5$) population --- including the carbon stars --- to be producing detectable amounts of dust. We compare the dust production to the stars' H$\alpha$ line profiles obtained
at the Magellan/Clay telescope at Las Campanas Observatory, finding pulsation shocking in the strongest pulsators (V6, V17 and V42), but evidence of outflow in all other stars. We conclude that the onset of dust production does not signify a fundamental change in the material leaving the star. Our data add to a growing body of evidence that metallic iron dominates dust production in metal-poor, oxygen-rich stars, but that dust is probably not the primary accelerant of winds in this mass-metallicity regime.
\end{abstract}



\section{Introduction}
\label{IntroSect}

\begin{figure*}
\centerline{\includegraphics[height=0.47\textwidth,angle=-90]{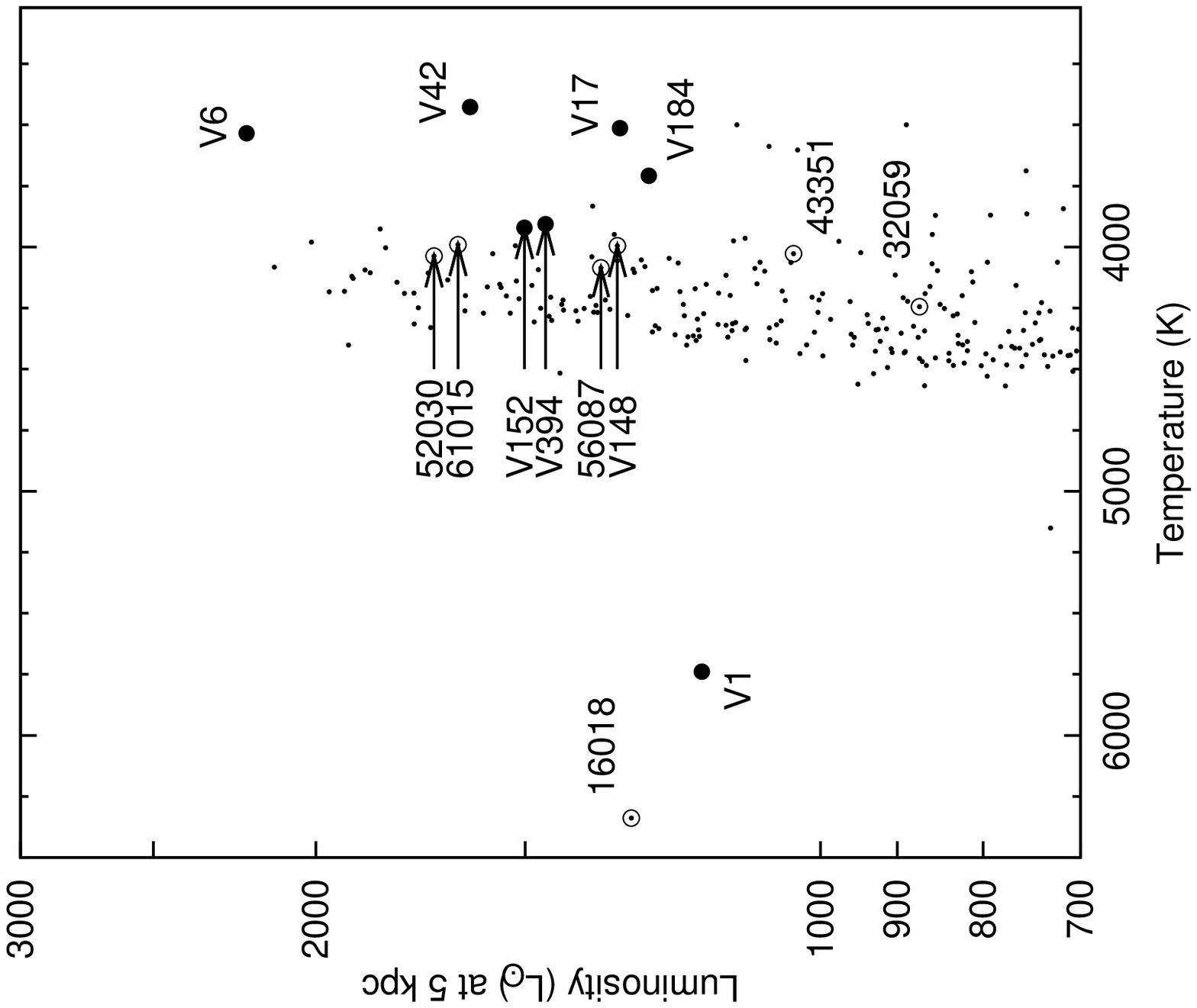}\includegraphics[height=0.47\textwidth,angle=-90]{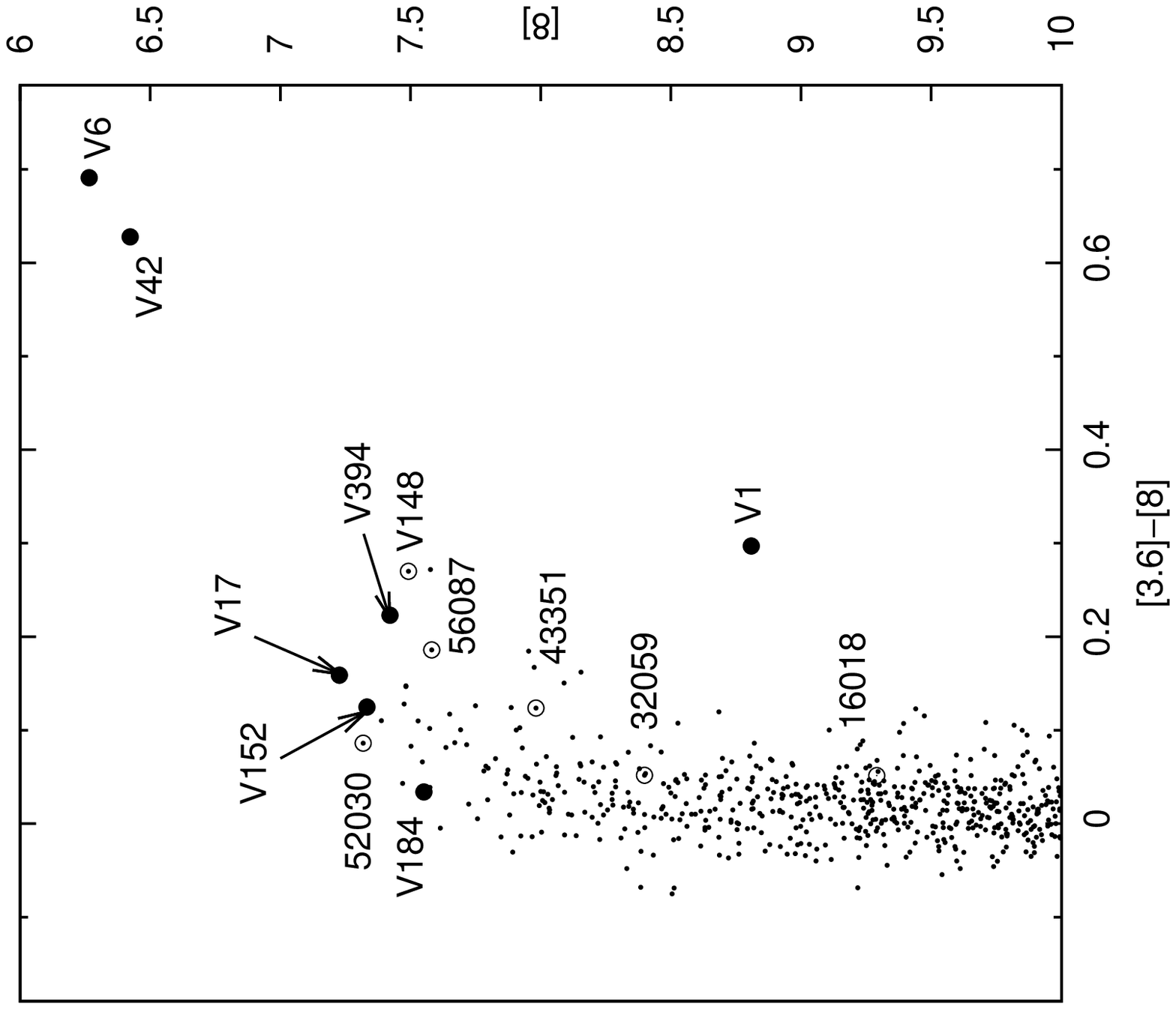}}
\caption{Hertzsprung--Russell and colour--magnitude diagrams showing the most-luminous giant stars in $\omega$ Cen from \protect\citet{MvLD+09} and \protect\citet{BMvL+08}, respectively. The large circles identify our targets, with filled squares identifying stars where dust is detected.}
\label{HRDFig}
\end{figure*}

\begin{figure*}
\centerline{\includegraphics[height=0.95\textwidth,angle=-90]{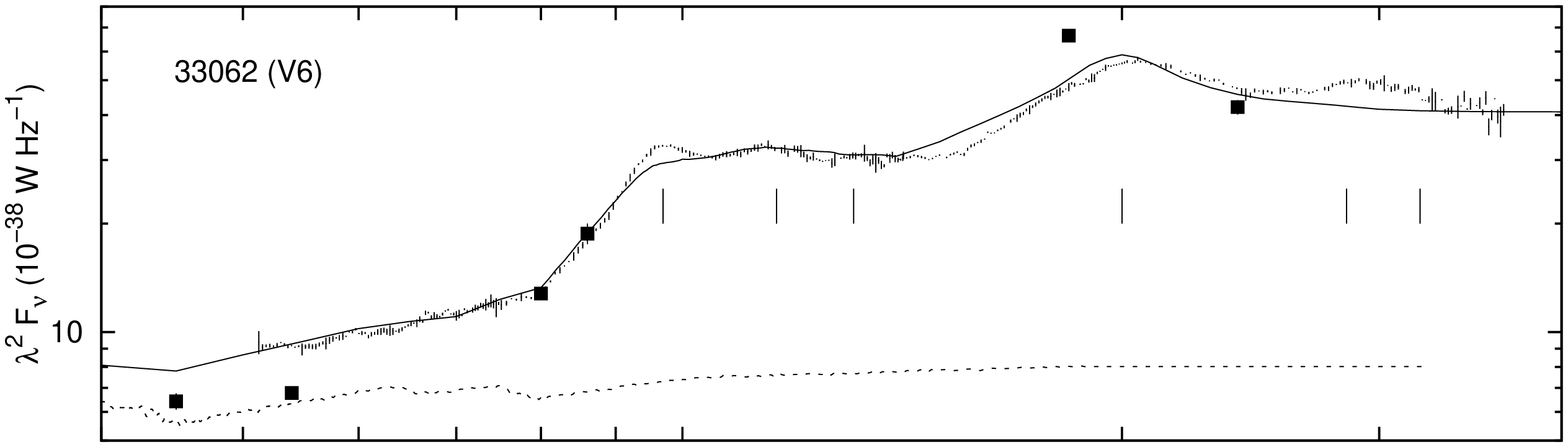}}
\centerline{\includegraphics[height=0.95\textwidth,angle=-90]{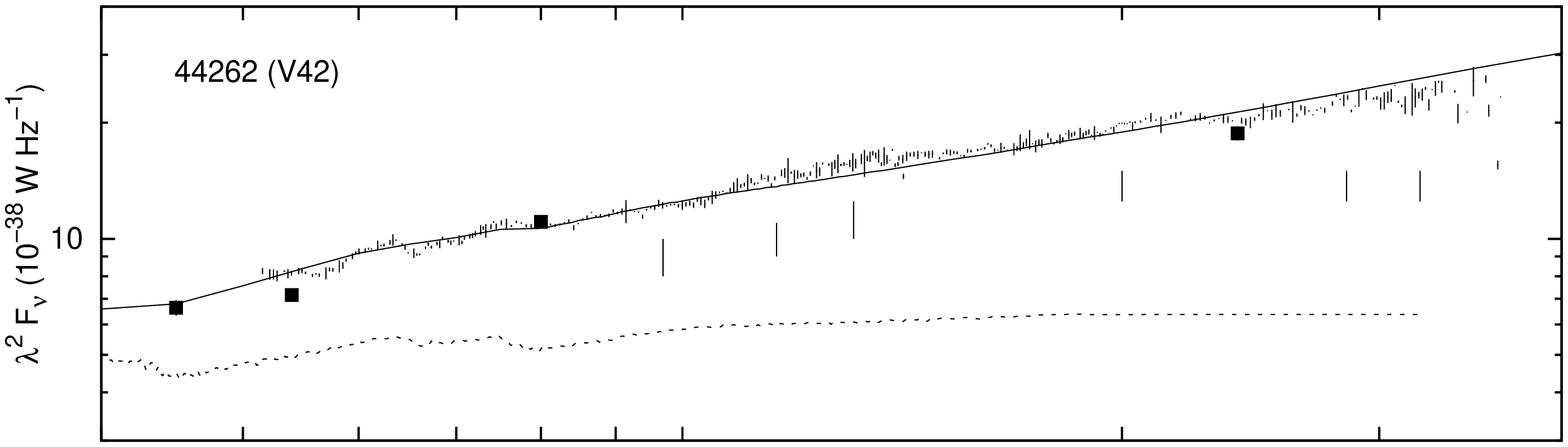}}
\centerline{\includegraphics[height=0.95\textwidth,angle=-90]{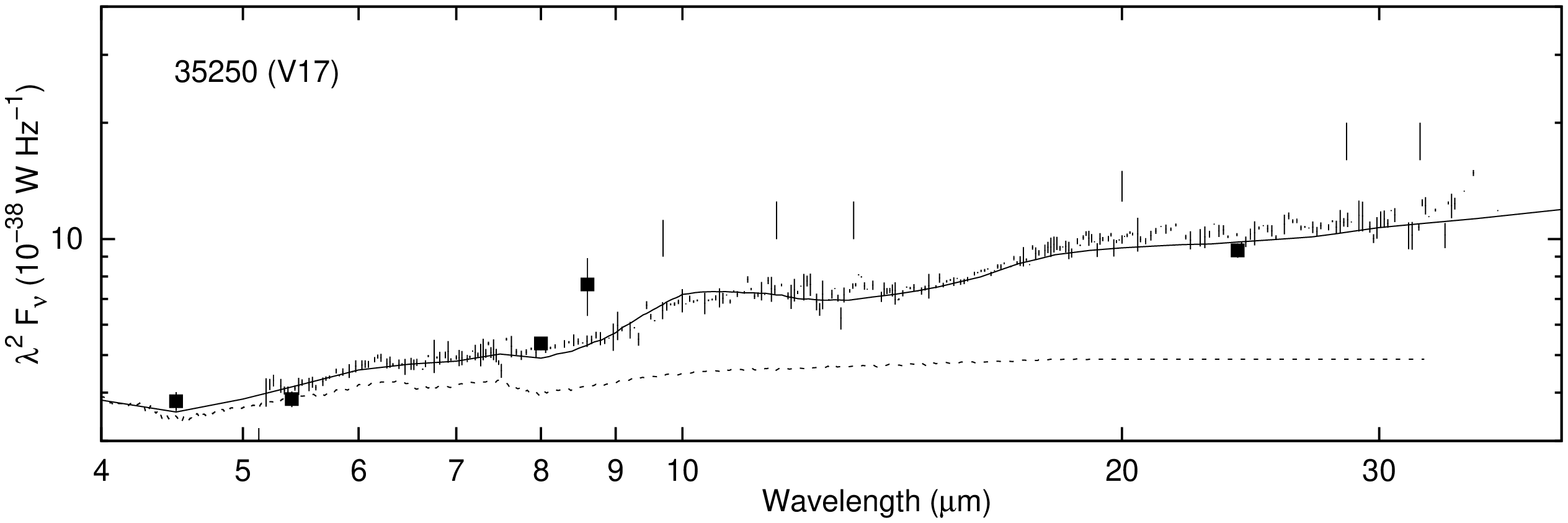}}
\vspace{10mm}
\caption{IRS spectra of V6, V42 and V17 (small points with error bars), showing the photospheric model fit (dashed line), literature photometric data (squares) and {\sc dusty} models (solid line). Vertical lines show the wavelengths of common crystalline features, which are not included in the {\sc dusty} models (\S\ref{V6Sect}). The stars are identified by their LEID.}
\label{IRSFig}
\end{figure*}

While the basis of how evolved stars eject mass is relatively-well understood (e.g.\ \citealt{Habing96}), accepted theory has difficulty in explaining mass loss from metal-poor stars. In the extended atmospheres of evolved stars, carbon (C) and oxygen (O) bond to form CO. The excess C or O forms C- or O-rich dust, generally amorphous carbon (AmC) or silicates, respectively. Radiation pressure on dust grains drives them from the star, and collisional coupling with the surrounding gas drives the gas with them \citep{GW71}. In metal-poor stars, however, there may be insufficient metals to both form dust grains and to maintain this gas--dust coupling. There may therefore be insufficient opacity to effectively drive the wind via radiation pressure alone: problems already exist in explaining dust driving in solar-metallicity winds \citep{Woitke06b}. It has been suggested that, at least in metal-poor stars, pulsation provides the primary acceleration that levitates material from the star \citep{vLCO+08}. Additional driving from magnetic or hydrodynamic processes in an active chromosphere or from dust driving will then merely act to modify the velocity of the outflow (e.g.\ \citealt{SD88,DSS09}).

The Galactic globular cluster $\omega$ Centauri provides an excellent laboratory for studying the evolution of dust production in metal-poor stars. As the most-massive ($2.5 \pm 0.3 \times 10^6$ M$_\odot$) and one of the closest ($\approx$5.0 kpc) globular clusters to Earth \citep{Harris96,vdVvdBVdZ06}, its large population of stars samples many different stages of evolution and dust production on the red and asymptotic giant branches (RGB, AGB). This population is split into a number of different metallicities: the bulk population with [Fe/H] $\approx$ --1.7, metal-intermediate populations of [Fe/H] $\approx$ --1.5 and --1.1, and an `anomalous' metal-rich population (the RGB-a) with [Fe/H] $\approx$ --0.8 \citep{Harris96,NFM96,PFB+00,SPF+05,VPK+07}. The leading hypothesis for the origin of these multiple populations is that $\omega$ Cen is the nucleus of a tidally-disrupted dwarf galaxy, possibly with a history similar to M54 in the Sagittarius dwarf spheroidal \citep{NB78,DKH79,Meylan87,Kamaya07}. Competing theories suggest multiple epochs of star formation in a self-enriching environment (see review: \citealt{Piotto09}). The cluster's high radial velocity ($+232$ km s$^{-1}$; \citealt{vdVvdBVdZ06}) allows for easy separation of cluster stars \citep{vLvLS+07}, while proper motions \citep{vLlPR+00} provide further membership confirmation. Recent studies (e.g.\ \citealt{JP10}; \citealt{MMP11}) also provide accurate metallicities and abundances for the majority of the cluster's most-luminous giant stars, allowing the effects of surface chemistry to be studied.

Mid-infrared (mid-IR) excesses caused by circumstellar dust have been found in several of $\omega$ Cen's stars. In particular, the stars V6 and V42 (LEID 33062 and 44262; \citealt{Clement97}; \citealt{vLlPR+00}) have been singled out as significant dust sources. V6 shows an excess redwards of 3.5 $\mu$m \citep{GF73,GF77} and silicate dust emission (\citealt{MvLD+09}, hereafter MvLD). V42 shows a similarly-strong mid-IR excess \citep{OFP95,OFFPR02,BMvL+08} and is O-rich (C/O $<$ 1; \citealt{MvL07}), but shows no silicate emission (MvLD). Despite the star's O-richness, MvLD attributed the lack of dust features in V42 to AmC. Under this hypothesis, AmC would form in non-equilibrium conditions in marginally oxygen-rich environments (see also \citealt{HA07}). Evidence has been found for AmC in CO novae, which are also thought to have oxygen-rich (C$<$O) ejecta, but this was based on the identification of other carbon-rich species not seen in these stars \citep{GTWS98}. The recent discovery that many stars in other clusters share this featureless mid-infrared excess \citep{vLMO+06,LPH+06,BMvL+09} led \citet{MSZ+10} to re-examine the phenomenon. They concluded that while the featureless excess emission could be caused by other materials, such as large silicate grains or AmC, these are not expected in metal-poor, oxygen-rich stars, meaning the most likely origin is pure metallic iron dust.

The existence of metallic iron dust as a star's sole dust species would be surprising for a number of reasons. Most globular cluster stars show super-solar [O/Fe] and [Si/Fe] ratios (e.g.\ \citealt{CBGL09,CBG+09}) and there is difficulty in explaining the optical depth with the amount of iron likely to be in it \citep{MBvLZ11}. Many dust condensation models predict that magnesium-rich silicates condense at high temperatures, with iron deposited in iron-rich silicates at lower temperatures (due to their higher opacity; e.g.\ \citealt{GS99}). Metallic iron is only expected to condense when atomic iron has high partial pressures in the wind, which should not be the case in metal-poor stars \citep{GS99,FG02,Woitke06b}. A full understanding of dust condensation around metal-poor stars must explain this. In this paper, we will investigate the dust production in $\omega$ Centauri, to see how it varies for a variety of targets at low metallicities (--2 $\lesssim$ [Fe/H] $\lesssim$ --1).

\section{Observations}

\subsection{\emph{Spitzer} IRS data}
\label{ObsSect}

\begin{center}
\begin{table*}
\caption{Targets observed with \emph{Spitzer} IRS, sorted by luminosity.}
\label{SourcesTable}
\begin{tabular}{l@{\ \ \ }l@{\ \ \ }lccccc@{\ \ \ }cc@{\ \ \ }cr@{\ \ \ }ll}
    \hline \hline
LEID$^1$	& ROA$^2$ & VID$^2$& Epoch$^3$& Period& Phase$^4$	& [Fe/H]& Temp.$^5$& Lum.$^5$& Temp.$^5$& Lum.$^5$ & \multicolumn{2}{c}{Excess$^6$}& Notes and\\
\ 		& \ 	& \ 	& (JD)	& (days)& 	&(dex)	& (K)	& L$_\odot$& (K)	& (L$_\odot$)	& \multicolumn{2}{c}{at 8 $\mu$m}	& references \\
    \hline
{\bf 33062}	&\nodata& V6	&27010.1&110: 	& 253.90&--1.08	& 3375	& 2278	& 3534	& 2199	& 76\%&	22$\sigma$	& 10,14,15\\
52030		& 55 	&\nodata&\nodata&\nodata&\nodata&--1.7  & 4022	& 1944	& 4037	& 1700	& 23\%&	7$\sigma$	& 7,9\\
61015		& 53 	&\nodata&\nodata&\nodata&\nodata&--1.71	& 4067	& 1881	& 3991	& 1645	& 11\%&	2$\sigma$	& 11\\
{\bf 44262}	& 9132 	& V42	&52525.7&149.4 	& 16.155&--0.8?	& 3565	& 1862	& 3427	& 1618	& 94\%&	27$\sigma$	& 14,16\\
{\bf 44277}	& 8187	& V152	&\nodata&124: 	&\nodata&--1.37	& 3963	& 1873	& 3921	& 1502	&\nodata&\nodata& 11,15,17\\
{\bf 55114}	& 132	& V394	&\nodata&\nodata&\nodata&--1.45	& 3996	& 1705	& 3906	& 1459	& 20\%&	5$\sigma$	& 11\\
56087		& 81	&\nodata&\nodata&\nodata&\nodata&--1.84	& 4209	& 1749	& 4085	& 1352	& 26\%&	7$\sigma$	& 11\\
41455		&\nodata& V148	&\nodata&90: 	&\nodata&--1.22	& 4111	& 1623	& 3995	& 1322	& 33\%&	9$\sigma$	& 11,15\\
{\bf 35250}	&\nodata& V17	&30062.2&64.725:& 384.36&--1.06	& 3606	& 1515	& 3513	& 1317	& 21\%&	6$\sigma$	& 10,14\\
16018		& 24 	&\nodata&\nodata&\nodata&\nodata&--1.91	& 6427	& 1480	& 6339	& 1297	& 3\%& 1$\sigma$	& 8,12\\
{\bf 42044}	& 320	& V184	&\nodata&0.30337&\nodata&--1.37	& 3780	& 1444	& 3708	& 1266	& 7\%& 2$\sigma$	& 10,15\\
{\bf 32029}	&\nodata& V1	&30027.0&29.3479& 848.85&--1.77	& 5675	& 1297	& 5739	& 1177	& 30\%&	9$\sigma$	& 13,14,15\\
43351		&\nodata&\nodata&\nodata&\nodata&\nodata&--0.98	& 4204	& 1258	& 4028	& 1038	& 12\%&	3$\sigma$	& 11\\
32059		& 134 	&\nodata&\nodata&\nodata&\nodata&\nodata& 3212	& 992	& 4245	& 873	& 3\%& 1$\sigma$	& 7\\
    \hline
\multicolumn{14}{p{0.95\textwidth}}{(1) Leiden Identifier \citep{vLlPR+00}, where numbers in bold typeface represent stars where dust is detected; (2) Royal Observatory \citep{Woolley66} and variable numbers \citep{Clement97,KOT+04}; (3) epoch of maximum optical light (Julian Date -- 2\,400\,000); (4) number of pulsations between epoch and \emph{Spitzer} IRS observation; (5) temperature and luminosity from MvLD (left columns) and this work (right columns); (6) flux in excess of the modelled photosphere at 8 $\mu$m, as a fraction of the model; (7) carbon star; (8) Fehrenbach's star. Metallicities from: (9) \citet{Mallia77}; (10) \citet{VWS02}; (11) \citet{JP10}; (12) \citet{CdSSR01}; (13) \citet{GW94}. Variability information (colons denote uncertain/irregular periods): (14) \citet{Clement97}; (15) \citet{KOT+04}; (16) \citet{SMM+10}. (17) no reliable 8-$\mu$m flux.}
\end{tabular}
\end{table*}
\end{center}

\noindent
Fourteen of $\omega$ Cen's most evolved stars (Table \ref{SourcesTable}, Fig.\ \ref{HRDFig}) were selected for observation with the InfraRed Spectrograph (IRS; \citealt{HRvC+04}) on board the \emph{Spitzer Space Telescope} \citep{WRL+04,GRW+07}. Programme sources (PID 50458; PI: J.\ Th.\ van Loon) were selected primarily for their IR excess (MvLD), though we also included the post-AGB stars LEID 16018 (Fehrenbach's star) and 32029 (V1), and the carbon stars LEID 32059 and 52030 to provide a more heterogeneous sample of dust producers \citep{vLvLS+07}.

These 14 sources were reanalysed using the spectral energy distribution fitter from MvLD. We used the same {\sc marcs} model atmospheres \citep{GBEN75,GEE+08}, and assumed a distance of 5 kpc, a reddening of $E(B-V) = 0.12$ mag and a mass of $M = 0.7$ M$_\odot$. In addition to the data used by MvLD, we included the $UBVR_cI_c$ data of \citet{PGFB07}, \citet{BPB+09} and \citet{SlCCdSC10}, plus mid-IR photometry from {\it AKARI}'s 9-$\mu$m band \citep{KAC+10} and from MvLD, and metallicities from \citet{JP10}. These revised data and assumptions slightly alter the derived stellar parameters with respect to MvLD (Table \ref{SourcesTable}). The computed 8-$\mu$m flux excesses have also changed, with the result that LEID 42044 no longer has a significant ($>$3$\sigma$) excess\footnote{LEIDs 16018, 32059 and 61015 were not reported to have excess by MvLD.}.

\begin{figure}
\centerline{\includegraphics[height=0.47\textwidth,angle=-90]{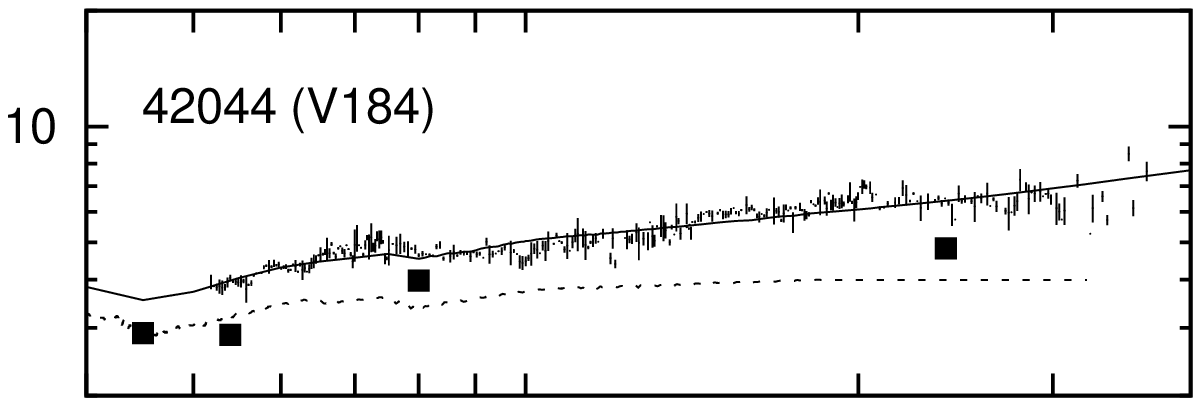}}
\centerline{\includegraphics[height=0.47\textwidth,angle=-90]{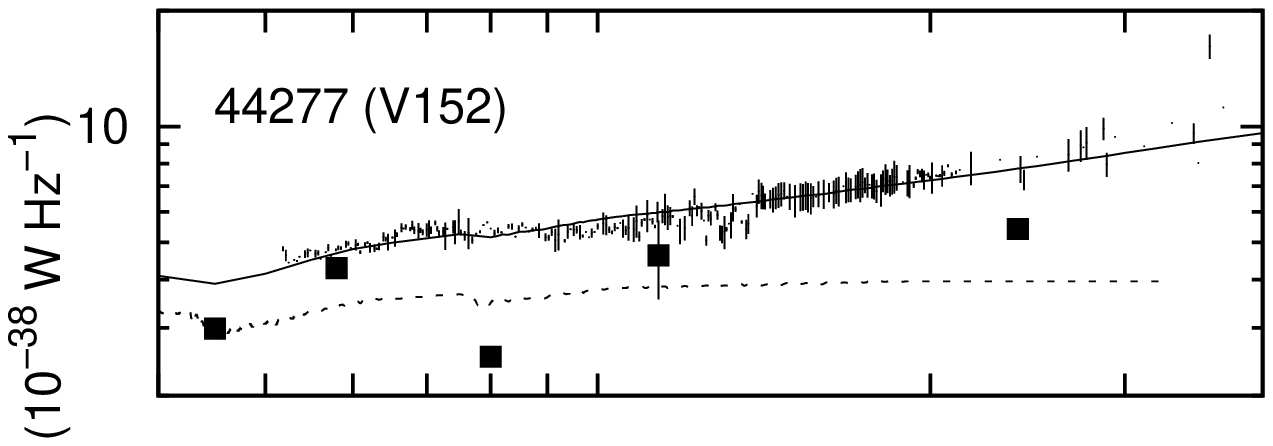}}
\centerline{\includegraphics[height=0.47\textwidth,angle=-90]{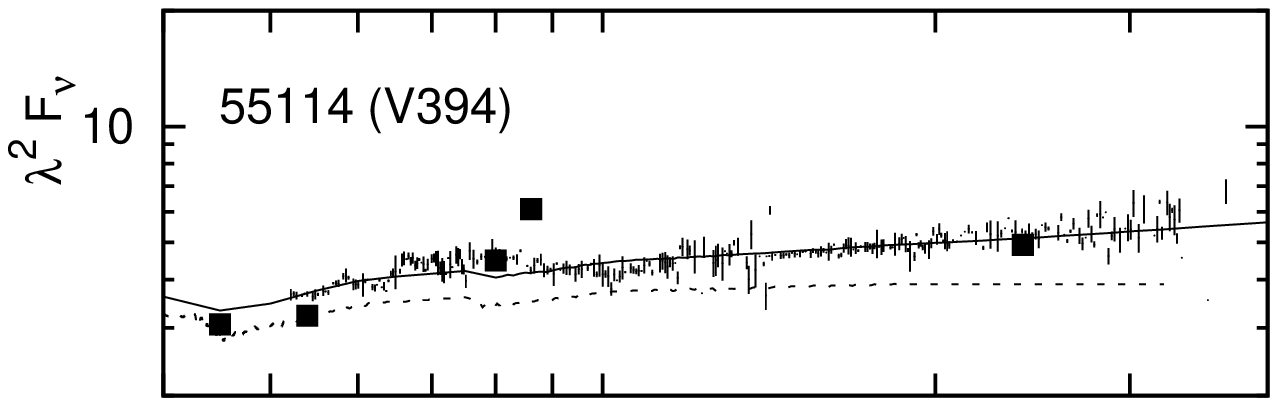}}
\centerline{\includegraphics[height=0.47\textwidth,angle=-90]{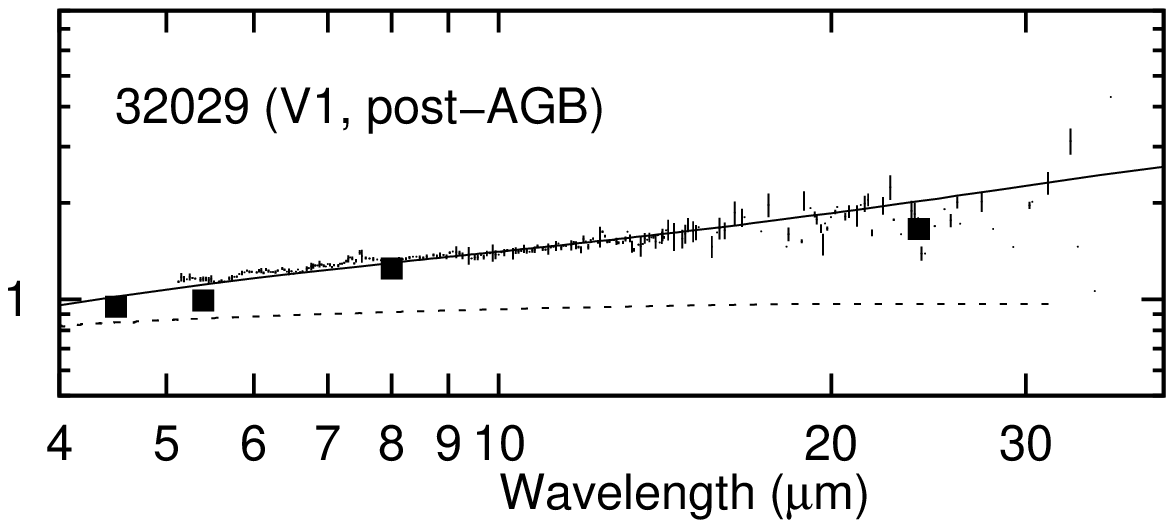}}
\vspace{7mm}
\caption{As Fig.\ \ref{IRSFig} for the other dusty stars.}
\label{IRS2Fig}
\end{figure}

\begin{figure}
\centerline{\includegraphics[height=0.47\textwidth,angle=-90]{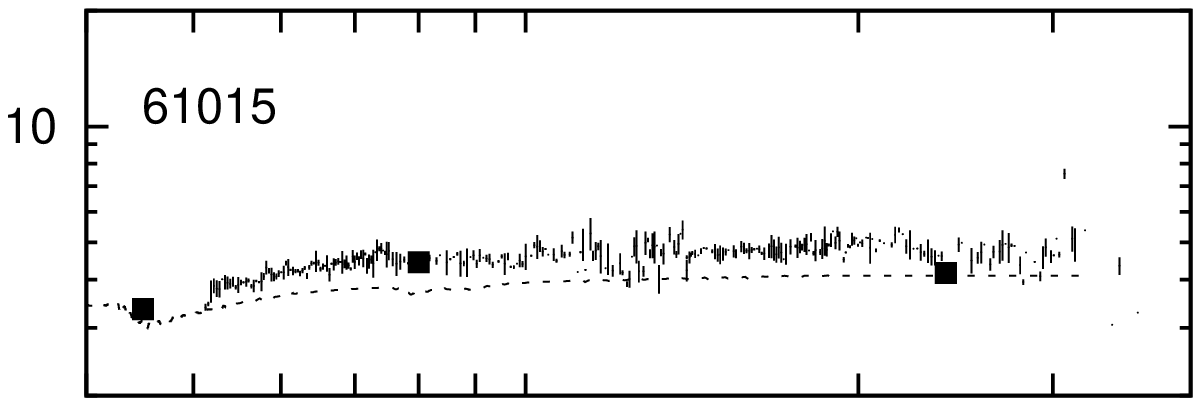}}
\centerline{\includegraphics[height=0.47\textwidth,angle=-90]{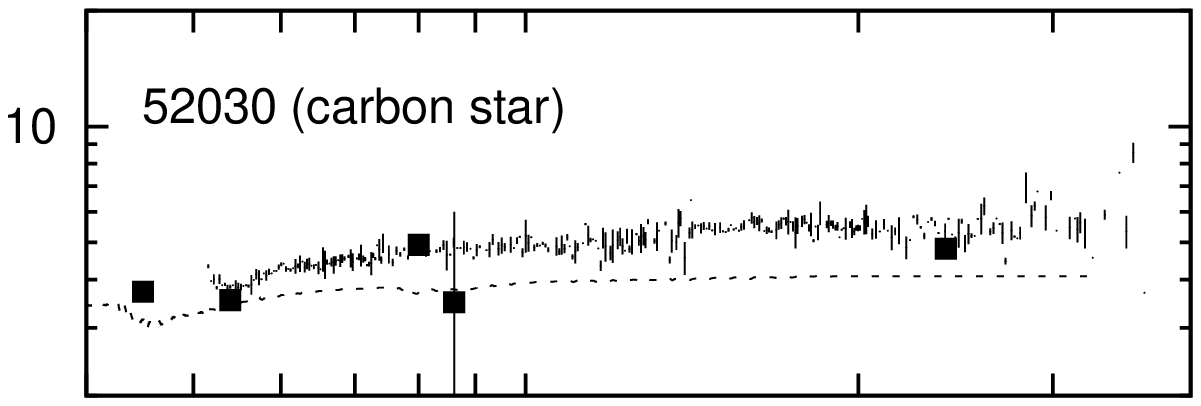}}
\centerline{\includegraphics[height=0.47\textwidth,angle=-90]{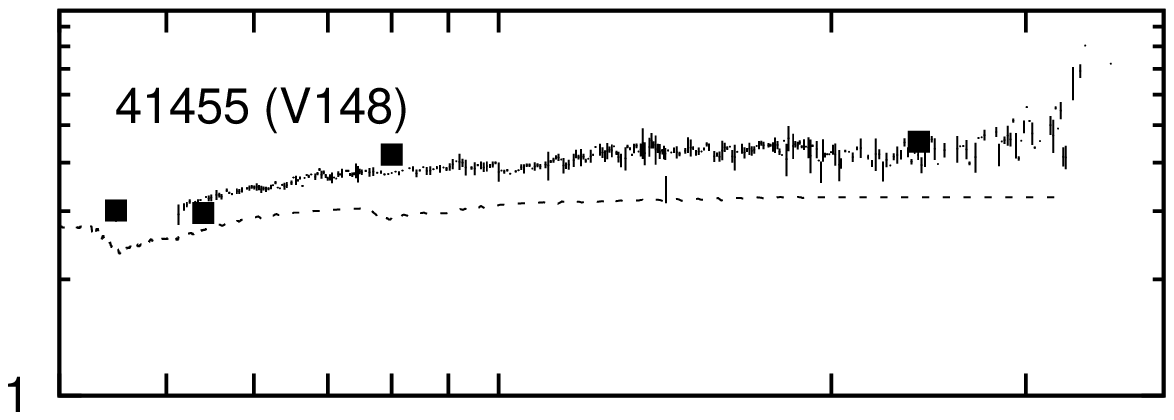}}
\centerline{\includegraphics[height=0.47\textwidth,angle=-90]{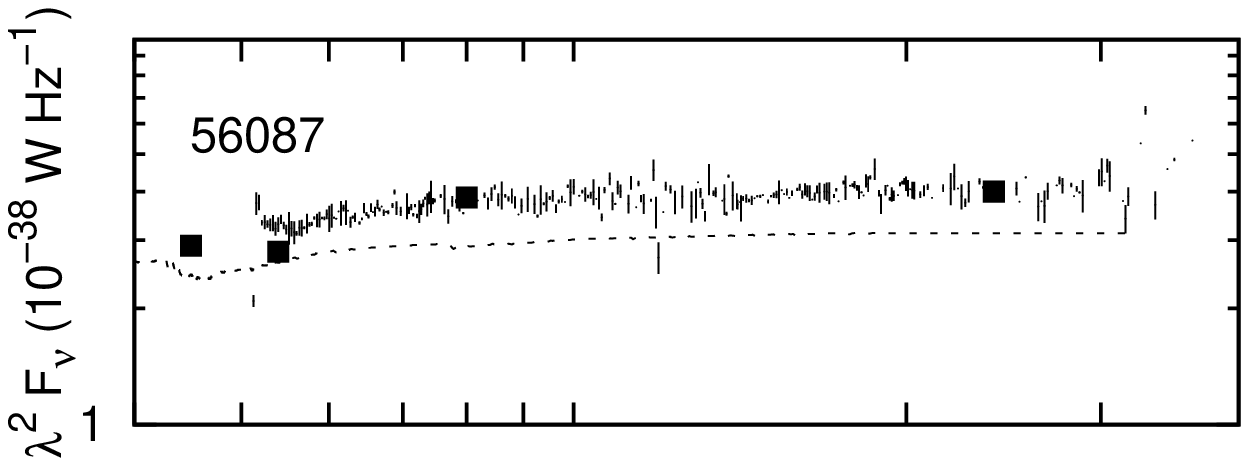}}
\centerline{\includegraphics[height=0.47\textwidth,angle=-90]{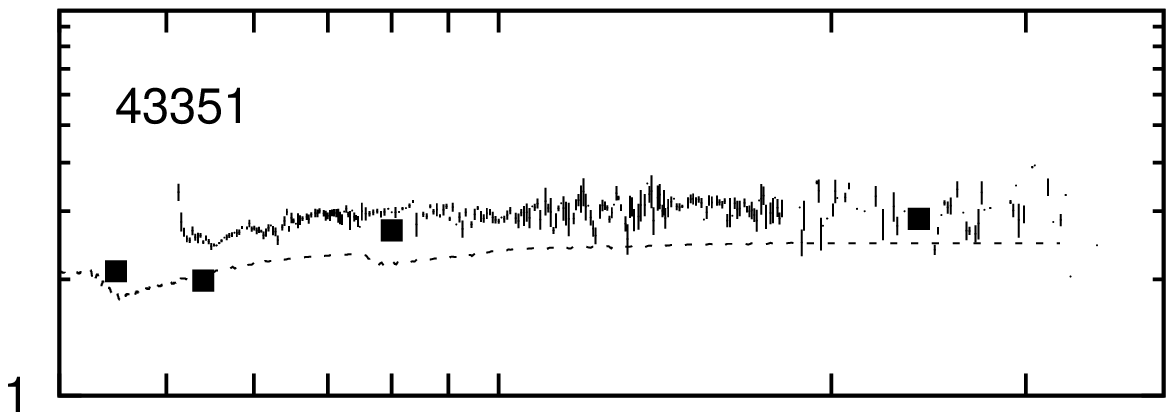}}
\centerline{\includegraphics[height=0.47\textwidth,angle=-90]{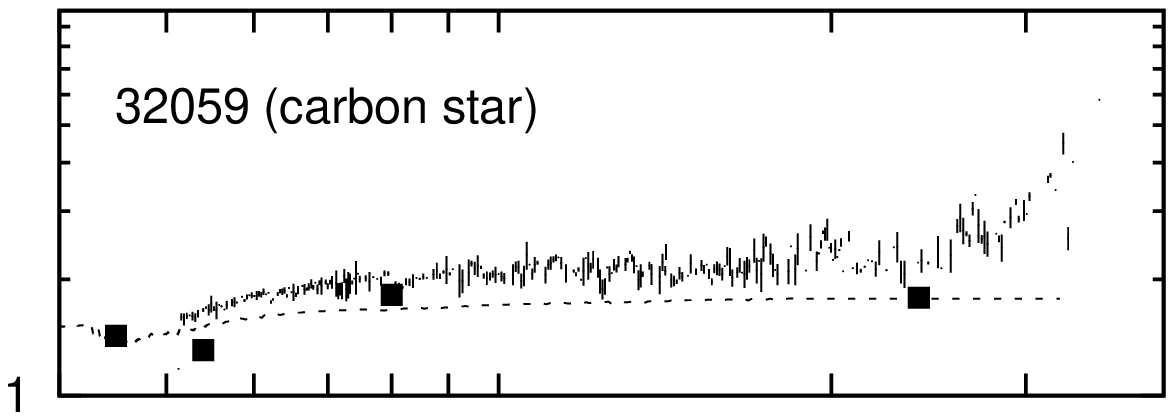}}
\centerline{\includegraphics[height=0.47\textwidth,angle=-90]{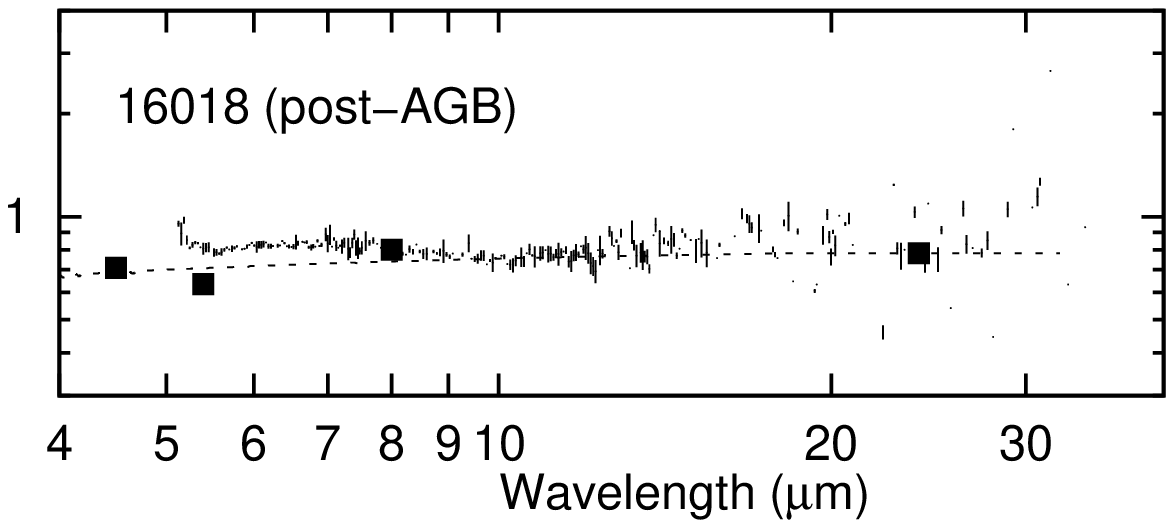}}
\vspace{7mm}
\caption{As Fig.\ \ref{IRSFig} for the dustless stars.}
\label{IRS3Fig}
\end{figure}

The extraction and calibration of these spectra followed the standard sequence (for details, see \citealt{SSM+10}). Spectra were extracted after images had been differenced to remove background radiation and cleaned to replace bad pixels. We extracted spectra from individual images, coadded them, and calibrated them with HR 6348 (for Short-Low; SL) and HR 6348 and HD 173511 (for Long-Low; LL).  The SL and LL spectra were joined, stitched, and trimmed. Stitching means that all segments were normalized multiplicatively (upward, by $\leq$14\%) to remove discontinuities between orders caused by pointing-induced errors in radiation passing through the spectroscopic slits. Trimming means removing extraneous data from the ends of the segments. The final spectra are shown in Figs.\ \ref{IRSFig}, \ref{IRS2Fig} \& \ref{IRS3Fig}, with flux plotted in $\lambda^2 F_\nu$, to better show deviations from a blackbody.

\subsection{Magellan H$\alpha$ spectra}
\label{HalphaObsSect}

We have obtained optical spectra of 12 of our 14 \emph{Spitzer} targets in order to compare gas mass loss tracers to the amount and type of dust produced. These spectra were taken with the Magellan Inamori Kyocera \'Echelle (MIKE) double-\'echelle spectograph mounted on the Magellan 6.5-m Clay telescope at Las Campanas observatory using a 0.75 $\times$ 5 arcsec$^2$ slit, giving a resolution of $R \approx 40\,000$. The spectra were reduced with bias subtraction, flat-fielding and sky subtraction using the MIKE-IDL pipeline updated by J.\ Prochaska\footnote{http://web.mit.edu/$\sim$burles/www/MIKE/}. Th--Ar arc exposures bracketting the stellar targets were used to determine the wavelength scale.

\begin{figure}
\centerline{\includegraphics[height=0.45\textwidth,angle=-90]{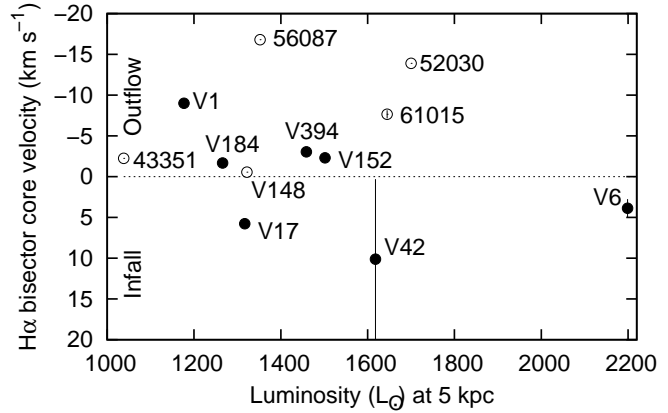}}
\caption{Bisector velocities of H$\alpha$ lines. Bars denote the range of velocities where multiple epochs are observed. Filled symbols show stars where dust is observed.}
\label{BisectFig}
\end{figure}

\begin{figure*}
\centerline{\includegraphics[width=0.88\textwidth,angle=0]{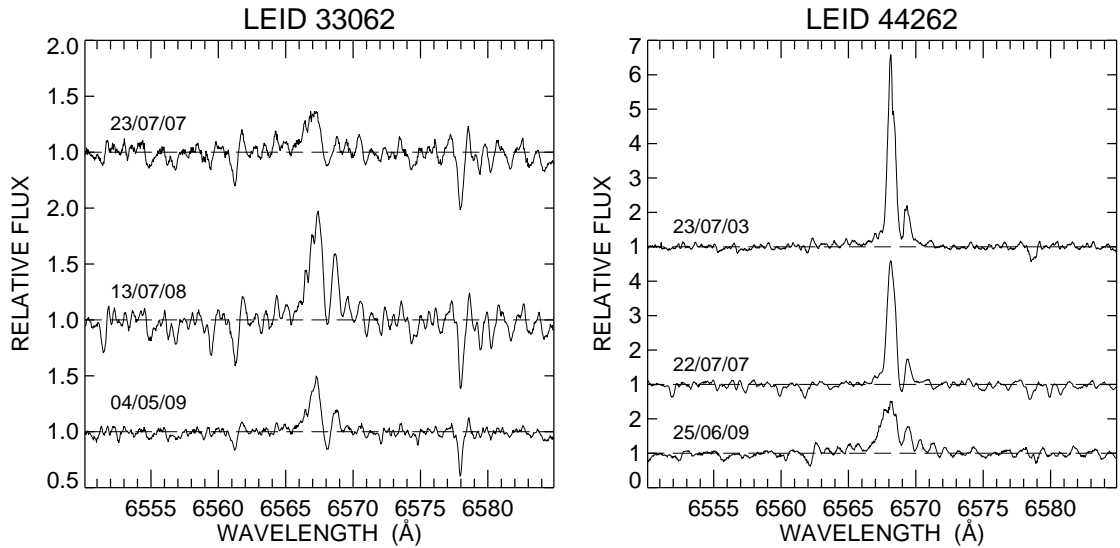}}
\caption{H$\alpha$ line profiles of LEID 33062 (V6) and LEID 44262 (V42).}
\label{HalphaFig2}
\end{figure*}

\begin{figure*}
\centerline{\includegraphics[width=0.88\textwidth,angle=0]{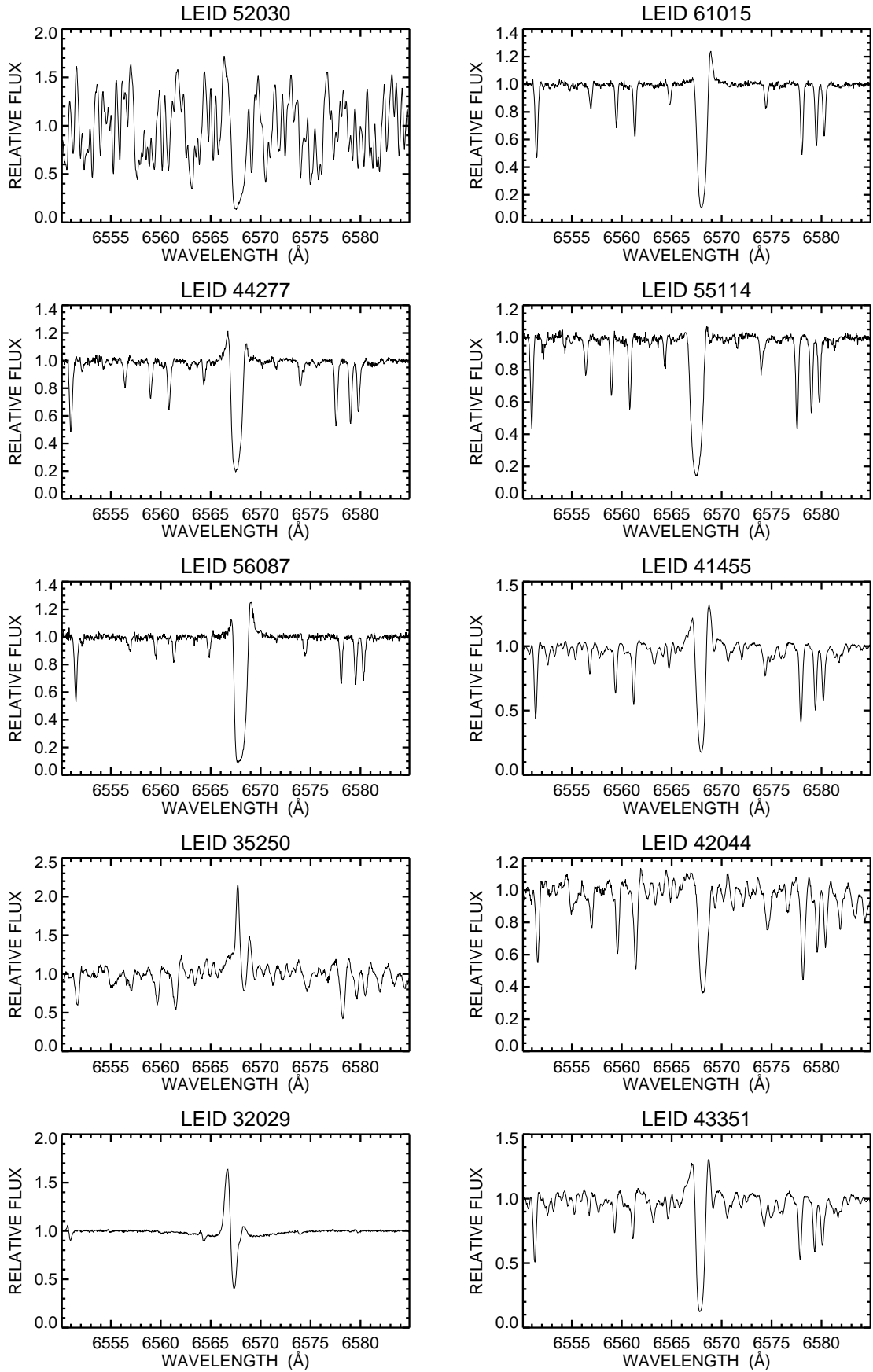}}
\caption{H$\alpha$ line profiles of programme targets, except LEID 33062 (V6) and LEID 44262 (V42), shown in Fig.\ \protect\ref{HalphaFig2}.}
\label{HalphaFig1}
\end{figure*}

H$\alpha$ line bisectors were determined and velocity shifts of these bisectors in the line cores were calculated with respect to stellar photospheric lines. These velocities generally trace the radial direction and strength of the gas flow in the $\sim$10\,000 K chromosphere at $\sim$1.4 to 1.8 $R_\ast$, but do not necessarily give an accurate outflow velocity and show instantaneous velocities for that radius. Modelling suggests actual outflow velocities are generally higher than bisector velocities derive, principally due to the superposition of emission and absorption components \citep{MCP06,MvL07,MAD09}. Radial and bisector velocities are listed in Table \ref{BisectTable} and shown in Fig.\ \ref{BisectFig}, with the original H$\alpha$ profiles shown in Figs.\ \ref{HalphaFig2} (V6 = LEID 33062 and V42 = LEID 44262) and \ref{HalphaFig1} (other targetted stars). The outflow velocity for the carbon star LEID 52030 was derived by comparing the position of the H$\alpha$ core to the centroid of the line at the continuum level. Typical errors associated with these velocities can be estimated from the range of bisector velocities among multiple spectra of the same target taken on the same night. These were found to be $\sim\pm$2 km s$^{-1}$.

\begin{center}
\begin{table}
\caption{H$\alpha$ bisector velocities for observed stars.}
\label{BisectTable}
\begin{tabular}{@{}l@{\ }l@{\ }c@{\ }c@{}c@{}c@{}l@{}}
    \hline \hline
LEID		& VID	& Observation	& Phase		& Heliocentric	& Bisector 	& Notes	\\
\ 		& \ 	& Date		& \ 		& radial	& velocity 	& \ 	\\
\ 		& \ 	& (UT) 		& \ 		& velocity	& \ 	 	& \ 	\\
\ 		& \ 	& \ 		& \ 		& (km s$^{-1}$)	& (km s$^{-1}$)	& \ 	\\
    \hline
{\bf 33062}	& V6	& 23/07/07	& 248.13	& 236.63	& +4.4		& Shocks\\
\ 		& \ 	& 13/07/08	& 251.38	& 238.91	& +2.6		& Shocks\\
\ 		& \ 	& 04/05/09	& 254.05	& 236.40	& +4.6		& Shocks\\
52030		&\nodata& 22/07/07	&\nodata	&\nodata	&--13.6$^1$	& Dense lines\\
61015		&\nodata& 23/07/07	&\nodata	& 238.77	& --7.7		& \\
{\bf 44262}	& V42	& 23/07/03	& 11.900	& 260.40	& +20.0		& Shocks\\
\ 		& \ 	& 22/07/07	& 11.900	& 260.40	& +20.0		& Shocks\\
\ 		& \ 	& 25/06/09	& 16.618	& 279.14	& +0.3		& Shocks\\
{\bf 44277}	& V152	& 22/07/07	&\nodata	& 217.68	& --2.7		& \\
\ 		& \ 	& 22/07/07	&\nodata	& 217.79	& --1.9		& \\
{\bf 55114}	& V394	& 24/07/07	&\nodata	& 217.11	& --3.0		& \\
56087		&\nodata& 24/07/07	&\nodata	& 240.65	& --16.8	& \\
41455		& V148	& 04/05/09	&\nodata	& 235.19	& --0.6		& \\
{\bf 35250}	& V17	& 24/07/07	& 374.56	& 247.60	&  +5.8		& \\
{\bf 42044}	& V184	& 23/07/07	&\nodata	& 244.53	& --1.7		& \\
{\bf 32029}	& V1	& 16/07/08	& 839.50	& 216.64	& --9.0		& Rev.\ P Cyg\\
43351		&\nodata& 05/05/09	&\nodata	& 230.94	& --2.2		& \\
    \hline
\multicolumn{7}{p{0.45\textwidth}}{$^1$Measured with respect to the line centroid, as dense molecular lines precluded a radial velocity measurement.}
\end{tabular}
\end{table}
\end{center}


\section{Modelling the dust features}
\label{ModellingSect}

Dust features were modelled using the radiative transfer code {\sc dusty} \citep{NIE99}, adopting optical constants from \citet{OBA+88} for metallic iron, \citet{DL84} for amorphous silicates, \citet{BDH+97} for amorphous alumina and \citet{HBMD95} for FeO. We assume the dust consists of solid spheres, while grain size populations follow the distribution of \citet{MRN77}. We further assume that, since the optical depth is low, the wind density follows {\sc dusty}'s numerical solution for radiatively-driven winds (density type = 3). A {\sc marcs} stellar atmosphere model was used for the underlying photosphere. This was interpolated from the grid used by MvLD using the temperature and metallicity listed in Table \ref{SourcesTable}.

Differences exist between the absolute flux calibration of the photometry in the literature (including \emph{Spitzer} IRAC/MIPS 3.6--24-$\mu$m fluxes from MvLD) and the IRS spectra. In general, the IRS spectra are brighter than the IRAC/MIPS photometry. This can occur due to source blending, and hence light from other sources entering the IRS module, or from intrinsic luminosity variations such as stellar pulsation. In the latter case, phase-corrected photometry can be used to eliminate this problem, but this cannot be done for irregular variables and/or those with unknown infrared pulsation amplitudes without contemporaneous infrared photometry. These mean that it can be difficult to determine whether an infrared excess is attributable to dust, or is an artifact. 

Source blending can usually be identified by an offset between the SL and LL components of the spectra, which observe different angular areas. The difference between the SL and LL fluxes can be used to approximate the photometric uncertainty: typically 5\%, and $\lesssim$17\% in all cases. Stellar pulsation, however, is more difficult to remove, especially in cases where pulsation periods are uncertain and/or variable. While the typical effect on infrared photometry may only be a few percent, strongly pulsating stars can vary by several tens of percent (see, e.g., \S\ref{V6Sect}), rendering the existence of an apparent small 8-$\mu$m excess in the single-epoch \emph{Spitzer} photometry meaningless. 

To complicate matters further, absorption or emission from the extended molecular atmosphere can also increase a star's infrared flux. The offending molecules in this case are predominantly H$_2$O and SiO. SiO produces an absorption band at 8 $\mu$m, which is included in the {\sc marcs} models, but the modelled equivalent width may vary from those of the stars themselves. We expect this may alter the 8-$\mu$m flux in the region of $\pm$2\% (a similar amplitude to the photometric error), but will have negligible effect on the continuum placement or the determination of the dust fraction.

Conversely, H$_2$O can provide absorption or emission throughout the spectrum and may affect, to some degree, the continuum placement and hence the amount and composition of dust. If this water layer is optically thick at infrared wavelengths, this emission will tend to approximate a single blackbody. This will multiplicatively increase the Rayleigh--Jeans tail of the stellar spectrum, resulting in an infrared spectrum that has a roughly constant value of $\lambda^2 F_\nu$, but is above that expected of the star's photospheric emission. When this extended envelope expands and forms optically-thin dust at larger radii, however, it will show a spectrum where $\lambda^2 F_\nu$ increases with $\lambda$. This criterion can be used to identify circumstellar dust in more ambiguous cases, and we only model stars where such a trend is clearly present. In the stars with weak metallic iron emission, there is thus a degeneracy between emission from circumstellar water and emission from the metallic iron dust. We estimate this may affect our dust production rates by a few $\times 10^{-11}$ M$_\odot$ yr$^{-1}$.

Of the 14 targets observed, only LEID 33062 (V6) and 35250 (V17) show 10- and 18-$\mu$m silicate dust features (weakly in the latter; \S\ref{V6Sect}). Several spectra vary as $F_\nu \propto \nu^{2}$, suggesting that the dust excess inferred from the photometry is not real and that these are naked (dustless) stars. A shallower (reddened) spectrum suggests the presence of an additional, cooler light source, namely circumstellar dust. Its featureless nature precludes its identification, but it can be assumed by process of elimination to be metallic iron (or AmC dust in the carbon stars), following \citet{MSZ+10}. This is seen in the O-rich stars LEIDs 32029 (V1), 35250 (V17), 44262 (V42), 44277 (V152) and possibly LEIDs 42044 (V184) and 55114 (V394). There is insufficient dust emission from LEIDs 16018, 41455 (V148), 43351, 56087 and 61015, and the carbon stars to differentiate them from naked stars. These stars may still exhibit infrared excess due to non-dusty components, such as molecular emission.

\begin{center}
\begin{table*}
\caption{Wind properties modelled using \protect{\sc dusty}.}
\label{DustyTable}
\begin{tabular}{llrrccccc}
    \hline \hline
LEID  & VID	& $\psi$ & $\tau_V$ & $T_{\rm cond}$ & $v_{\infty}$  & $v_{\rm esc}$ & $\dot{D}$  & $\dot{M}$ \\
\ 	& \ 	& \      & \        & (K)            & \llap{(}km s$^{-1}$\rlap{)} & \llap{(}km s$^{-1}$\rlap{)} & \llap{(}M$_\odot$ yr$^{-1}$\rlap{)} & \llap{(}M$_\odot$ yr$^{-1}$\rlap{)} \\
    \hline
 33062$^1$&  V6	&  2400  & 0.30     & 900            & 2.06          & 44.6	& $7\times10^{-10}$     & $2\times10^{-6}$ \\ 
 44262	&   V42	&  1260  & 0.58     & 950            & 2.97          & 46.7	& $1\times10^{-9}$      & $2\times10^{-6}$ \\ 
 44277	&  V152	&  4690  & 0.13     & 1050           & 1.43          & 54.4	& $2\times10^{-10}$     & $1\times10^{-6}$ \\ 
 55114	&  V394	&  5640  & $\sim$0.03&1000           & 0.87          & 54.6	& $6\times10^{-11}$     & $4\times10^{-7}$ \\ 
 35250$^2$& V17	&  2300  & 0.07     & 700            & 0.90          & 49.4	& $3\times10^{-10}$     & $6\times10^{-7}$ \\ 
 42044	&  V184	&  4690  & $\sim$0.02&700            & 0.49          & 53.7	& $8\times10^{-11}$     & $4\times10^{-7}$ \\ 
 32029	&    V1	& 11780  & 0.05     & 1100           & 0.86          & 84.7	& $7\times10^{-11}$     & $8\times10^{-7}$ \\ 
    \hline
\multicolumn{9}{p{0.7\textwidth}}{Models include (1) 70\% iron, 17\% silicates, 8\% Al$_2$O$_3$, 5\% FeO; and (2) 92\% iron, 8\% silicates by number. All other models are for 100\% iron grains (see text). The escape velocity ($v_{\rm esc}$) is determined at 2 R$_\ast$ assuming a stellar mass of 0.65 M$_\odot$.}\\
\end{tabular}
\end{table*}
\end{center}

Table \ref{DustyTable} summarises our model fitting results, including the assumed dust-to-gas ratio ($\psi = 200 \times 10^{\rm -[Fe/H]}$), $V$-band optical depth ($\tau_V$), dust condensation temperature ($T_{\rm cond}$), terminal velocity ($v_{\infty}$) and mass-loss rate for dust only ($\dot{D}$) and dust plus gas ($\dot{M}$). Mass-loss rates and velocities are calculated from {\sc dusty} output using the parameterisation \citep{NIE99}:
\begin{eqnarray}
 \dot{M} &=& \dot{M}_{\rm DUSTY} \left(\frac{L}{10^4\ {\rm L_\odot}}\right)^{3/4} \left(\frac{\psi}{200} \frac{\rho_{\rm s}}{3\ {\rm g cm^{-3}}}\right)^{1/2} ,\\
 \dot{M} &=& \dot{D} \psi ,\\
 v_{\infty} &=& v_{\rm DUSTY} \left(\frac{L}{10^4\ {\rm L_\odot}}\right)^{1/4} \left(\frac{\psi}{200} \ \frac{\rho_{\rm s}}{3\ {\rm g cm^{-3}}}\right)^{-1/2} .
\label{VEqun}
\end{eqnarray}
For modelling purposes, we assume that dust grains are solid, i.e.\ that they have grain densities ($\rho_{\rm s}$) of 7.9 g\,cm$^{-3}$ for metallic iron, and $\approx$3 g\,cm$^{-3}$ for other condensates.

We stress that the values in Table \ref{DustyTable} depend strongly on the dominant wind-acceleration mechanism. We assume here that grains start from a velocity of zero where the grains condense and are accelerated by radiation pressure from the star. The velocities listed in Table \ref{DustyTable} are exceptionally low: any additional `kick' velocity a grain receives, either during or after its formation, will increase the derived $\dot{M}$ proportionally. Uncertainties in grain size, and thus surface-area-to-volume ratio, add further uncertainty to the wind velocity, though $\dot{M}$ is hardly affected. The values in Table \ref{DustyTable} can therefore be thought of as highly uncertain, particularly $v_\infty$, but serve as a relative comparison between stars. The fractional compositions listed in Table \ref{DustyTable} can be determined quite precisely from the {\sc dusty} fits, but their absolute accuracy also depends very much on the above grain properties (see \S\ref{V6Sect}).

\section{Discussion}
\label{DiscSect}

\subsection{Correlating gas and dust outflows}
\label{BiSect}

A direct correlation between the H$\alpha$ outflow velocity, the presence of dust and the dust velocity is not necessarily expected: modification of the photospheric H$\alpha$ line occurs in the chromosphere near the star ($\lesssim$2 R$_\ast$), with the emission in the line wings produced close to the star and additional absorption in the line core produced slightly further out \citep{MAD09}. Dust, on the other hand, is produced in the cooler expanding envelope at $\gtrsim$2 R$_\ast$ (e.g.\ \citealt{WQ08}).

In the most-evolved stars, strong pulsation may augment chromospheric heating, with bright emission components coming from compression- and/or shock-heated gas, some of which may lie above the H$\alpha$ core forming region. Stars exhibiting these phenomena will show strong emission in the wings of the Balmer line, with much shallower line cores. LEID 33062 (V6) and 44262 (V42) show such emission in all Balmer lines\footnote{H$\epsilon$ is obscured by the Ca {\sc ii} K line \citep{vLvLS+07}. See \citet{MvL07} for another H$\alpha$ profile of V42.}, while LEID 35250 (V17) shows such emission in H$\alpha$, but not in the other Balmer lines (\citealt{vLvLS+07}; non-contemporaneous). Apparent infall suggested in these stars (Fig.\ \ref{BisectFig}) likely arises from the shock dynamics or other compression-related departures from LTE.

The other nine stars with H$\alpha$ observations show strong absorption in all Balmer lines, where they are observed \citep{vLvLS+07}. This shows that the compression heating present in the strongest pulsators (V6, V17 and V42) is not important in these stars, meaning the H$\alpha$ outflow velocities are easier to interpret. With the exclusion of these three stars, the remaining stars all show blueshifted H$\alpha$ line cores, indicative of outflow (though the outflow of V148 is also consistent with zero outflow).

The lack of strong emission due to shock excitation in the Balmer lines of the five faintest dust producers (V1, V17, V184, V394 and V152; \citealt{vLvLS+07}) suggests that substantial compression is not required for dust production to take place. The existence of such compression in the heavily dust-enshrouded V6 and V42, however, suggests that it may cause more-effective dust production.

Gas outflow, as traced by H$\alpha$ line core velocities, appears present in all the stars for which it is a valid tracer. The ubiquity of this outflow confirms that mass loss occurs among all stars near the RGB and AGB tips, regardless of the existence of circumstellar dust. Dust production therefore depends on mass loss, and not the other way round. The onset of dust production may therefore only be a byproduct of crossing a fundamental threshold in wind conditions, rather than a change in the amount of mass lost or the method by which the wind is driven from the star.

\subsection{Silicate emission in V6 and V17}
\label{V6Sect}

LEID 33062 (V6) is the only star showing strong solid-state emission features in the infrared. The dust excess includes a sharp 9.6-$\mu$m feature with a shoulder at 11.4~$\mu$m, a broad feature at $\sim$20~$\mu$m (the only feature successfully modelled here) and, possibly, a low-contrast feature at $\sim$13~$\mu$m. The spectral structure at 9--12~$\mu$m is most consistent with a mixture of amorphous silicates and crystalline forsterite (Mg$_2$SiO$_4$; see \citealt{SDBS+06}, Fig.\ 5). In this spectrum, the usual 18-$\mu$m feature is shifted to closer to 20~$\mu$m, likely due to an additional low-contrast component centered at $\sim$20~$\mu$m, but somewhat less pronounced and sharp than often seen (e.g.\ \citealt{SKGP03}, Fig.\ 1; \citealt{LPH+06}, Fig.\ 2). The dust in LEID 33062 is consistent with the previously noted correlation between the strength of the features at 13 and 20~$\mu$m \citep{SKGP03}. In LEID 33062, the 20~$\mu$m contribution is present, but relatively weak, and the 13~$\mu$m feature, while not definitely present, is weaker still.

There are two interpretations of these additional components (see \citealt{SMM+10} and references therein). They might arise from a combination of spinel (MgAl$_2$O$_4$, 13~$\mu$m) and magnesio-w\"{u}stite (Mg$_x$Fe$_{1-x}$O, 20~$\mu$m), but they might arise instead from ``warm crystalline dust''. In the latter scenario, crystalline alumina (Al$_2$O$_3$) produces the 13~$\mu$m feature, while a crystalline silicate such as enstatite (MgSiO$_3$) produces the 20-$\mu$m component. This would point to increased annealing of the dust grains already present (e.g.\ by shock compression), while the former scenario points to modified chemistry in the outflows.

LEID 33062 joins a growing sample of evolved globular cluster stars showing these warm crystalline dust features. In globular clusters, these features are seen in 47 Tuc V4, V8, V13, and V21 (\citealt{LPH+06,MBvLZ11}), NGC 5927 V1, and NGC 6352 V5 \citep{SMM+10,MSZ+10}. In the LMC, HV 2310 and HV 12667 also show features between 9 and 12 $\mu$m \citep{SKW+08} but these more massive and metal-richer stars may show different dust features to the globular cluster stars in the 13--14.5-$\mu$m region.

V17 (LEID 35250) shows evidence of silicate emission at 10 and 20 $\mu$m, but the silicate emission is much weaker than that of V6. The noisier, lower-contrast spectrum of V17 prevents us from confidently identifying the presence of crystalline silicate features. Several small deviations in the spectrum hint at their existence, however (Fig.\ \ref{IRSFig}). The most promising of these is the 13-$\mu$m feature.

A homogeneous model fails to reproduce the 10- and 18-$\mu$m amorphous silicate features of both stars simultaneously with the iron: the 18-$\mu$m feature is usually modelled to be too bright. One possible reason is that it is due to a difference in temperature between the silicate and iron grains, as discussed by \citet{MSZ+10}. The fraction of silicates modelled for V6 and V17 is also very small (17\%--22\% and 5\%), compared to the metallic iron (70\% and 95\%). This could be due to differences in grain size/shape distribution between the two grain populations (this has also been suggested in 47 Tuc; \citealt{MBvLZ11}).

\subsection{Iron emission and the metallicity of V42}
\label{V42Sect}

\begin{figure*}
\includegraphics[height=0.95\textwidth,angle=-90]{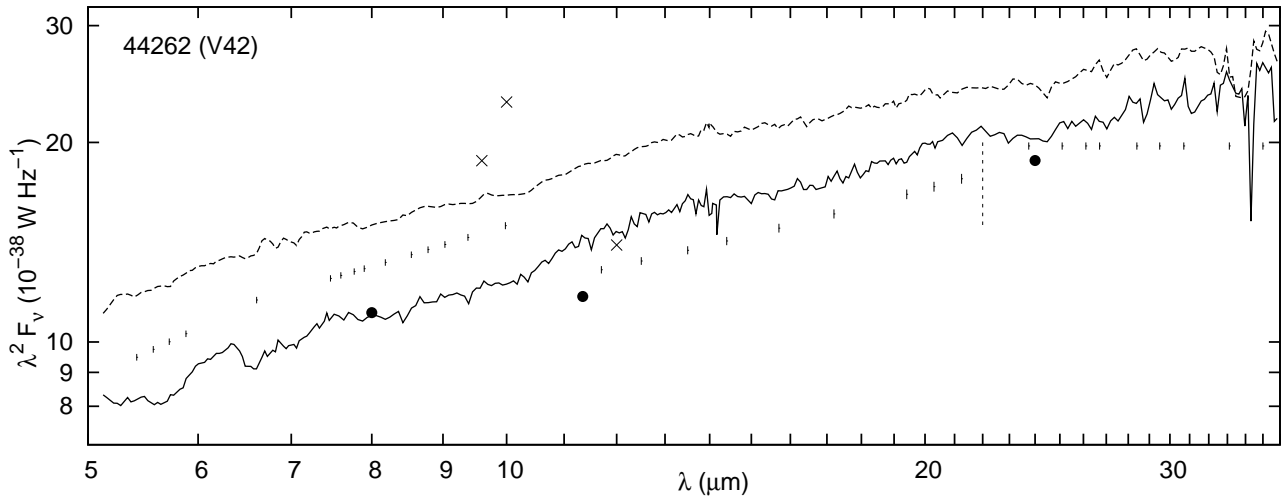}
\caption{Variations in the spectrum of V42 between this work (solid line) and \citet{SMM+10} (dashed line). The vertical dashes show water lines in absorption ($\lambda < 22 \mu$m) and emission ($\lambda > 22 \mu$m). Line positions were obtained from SpectraFactory \citep{CvMM10}. Literature photometric data, taken at various pulsation phases, are also shown (crosses: \citealt{OFFP95}; \citealt{OFFPR02}; large points: \citealt{BMvL+08}; MvLD).}
\label{V42Fig}
\end{figure*}

\begin{figure*}
\includegraphics[height=0.95\textwidth,angle=-90]{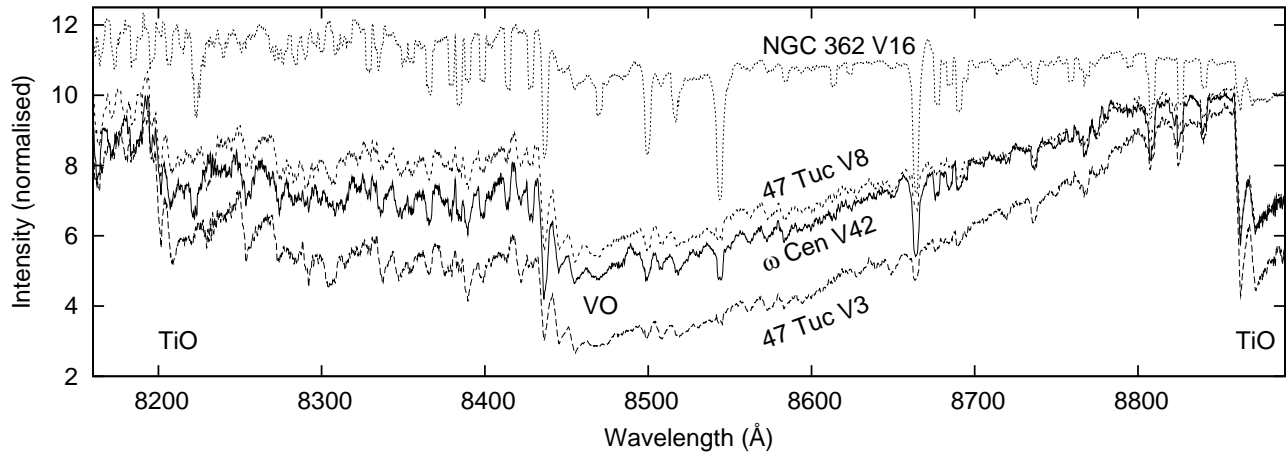}
\caption{UVES spectra from \protect\cite{MvL07}, smoothed with a Lorentzian profile to $R \sim 1000$. The stars shown are (bottom to top) 47 Tuc V3, $\omega$ Cen V42, 47 Tuc V8 and NGC 362 V16. V3 and V8 have the same metallicity ([Fe/H] = --0.7) but differ in temperature by $\sim$400 K. V16 is more metal-poor ([Fe/H] = --1.2) but is another $\sim$400 K warmer than V8.}
\label{UVESFig}
\end{figure*}

V42 (LEID 44262) had already been observed by \emph{Spitzer} \citep{SMM+10}, but was re-observed to check for evidence of a variable 10-$\mu$m emission feature (MvLD). The two spectra were taken on MJD 54\,310.7 and 54\,939.2, corresponding to pulsation phases 11.948 and 16.155 (epoch from MvLD). The absolute flux of the spectra is markedly different due to changes in the incident flux from the central star. The shape of the spectra, however, is virtually identical, except for a slight change in slope (Fig.\ \ref{V42Fig}). This suggests that the composition of circumstellar dust does not alter significantly over the pulsation cycle, though the dust temperature may. Significant differences do occur in V42's water lines around 6--7 $\mu$m, with some lines transitioning from emission to absorption. This implies that excitation of the water molecules is more strongly affected by pulsation than dust formation is.

The change in V42's dust temperature and luminosity without significant change in its dust chemistry is concurrent with similar observations of T Cep (Niyogi, Speck \& Onaka 2011), and is likely due to changes in irradiation caused by the pulsation of the central star. V42's suspected 10-$\mu$m emission feature could be due to both systematic over-estimation of photometry due to the crowded observation field and changes in the star's intrinsic brightness over the pulsation phase.

V42 is also one of the few bright giants in the cluster without an accurate metallicity determination. Such determinations are difficult due to the star's strong molecular bands. Previous spectroscopic surveys of the cluster have failed to determine its metallicity: \citet{vLvLS+07} find it may belong to the metal-enhanced populations ([Fe/H] = --1.25) but their metallicity determinations for individual stars are hampered by the comparatively low resolution of their $B$-band spectra.

The temperature of the star itself may provide the best clue to its metallicity. Metal-poor stars, due to their lower atmospheric opacity, have hotter temperatures at a given phase of evolution. A circumstellar extinction of $\tau_V = 0.58$ mag, translates to an apparent cooling of V42 by $\sim$300 K. Although approximate, raising the observed effective temperature of V42 by this amount effectively removes the circumstellar reddening, revealing the photospheric temperature. This still leaves it considerably cooler than the cluster's giant branch (cf.\ Fig.\ \ref{HRDFig}), placing it within the RGB-a sequence, suggesting it has a similar metallicity of [Fe/H] $\sim$ --0.8.

Comparison with spectra of stars from more metal-rich globular clusters (Fig.\ \ref{UVESFig}; spectra from \citealt{MvL07}) indicates that V42's spectrum is similar to those of the most-evolved stars in 47 Tuc ([Fe/H] = --0.7). The other comparison star in Fig.\ \ref{UVESFig} (NGC 362 V16) has similar pulsation properties and shows a similar infra-red spectrum to V42 (both in terms of dust composition and amount of excess; \citealt{SMM+10}), and is likely at a similar evolutionary stage. V16, however, is a factor of $\sim$3 more metal-poor at [Fe/H] $\sim$ --1.2. As the most evolved star in NGC 362, V16 should be at a similar, or more-evolved, evolutionary stage compared to V42. Despite this, Fig.\ \ref{UVESFig} shows NGC 362 V16 is clearly warmer than V42, with much less opaque TiO and VO bands, suggesting V42 has a metallicity more in line with 47 Tuc's [Fe/H] = --0.7.

As an additional test, we compared V42's spectrum with the synthetic spectral library provided by Coelho et al. (2005). We investigated the
parameter space $T_{\rm eff}$ = 3500 K and [Fe/H] = --2.5 to +0.5, in 0.5 dex increments. A general lack of spectral models below 3500 K makes an accurate metallicity determination difficult. From these comparisons, however, we concluded that V42 likely lies between [Fe/H] = --1 and [Fe/H] = 0.

It would appear, {\it prima facie}, that V42 belongs to the anomalous RGB, at [Fe/H] $\approx$ --0.8. However, while we consider it unlikely, we cannot rule out at the present stage that it may come from $\omega$ Cen's metal-intermediate ([Fe/H] = --1.5, --1.2, --1.0) populations \citep{JP10}. This determination decreases the derived mass-loss rate and increases the terminal wind velocity compared to a star with average metallicity ([Fe/H] = --1.62) by a factor of 2.5$\times$, yielding the values listed in Table \ref{DustyTable}.

\subsection{Global dust production in $\omega$ Cen}
\label{ProgressSect}

Our target selection sampled a diverse range of stellar parameters, such as evolution, surface chemistry, metallicity and luminosity. Though our sample is by no means unbiased (see below), we obtain a much more representative sample of a globular cluster's dust production than previous studies. \citet{LPH+06}, the closest similar study, focussed on the higher-metallicity ([Fe/H]$\sim$--0.7) cluster 47 Tuc and observed stars chosen for their luminosity and pulsation modes, resulting in their more homogeneous sample. From our diverse sample, we arrive at some significant conclusions:

(1) Dust production in the cluster is limited to a small number of highly-evolved objects. Several of our 14 targets have no dust emission,  corroborating our recent finding that dust production first occurs in stars above 1000 L$_\odot$ and becomes ubiquitous at $\sim$2000 L$_\odot$ (MvLD; \citealt{BMvL+09}; \citealt{MBvL+11}; \citealt{MBvLZ11}). Stellar temperature would appear to be the main factor influencing whether dust is produced or not: with the exception of the post-AGB star V1, dust-producing stars all have temperatures of $\lesssim$3950 K, while dustless stars are all $\gtrsim$3950 K. However, an intentional selection bias was introduced to favour cooler targets that may have been reddened by dust.

(2) Despite being few in number, the metal-rich population appears to produce most of the dust. Assuming V42 is metal-rich, the post-AGB star V1 (LEID 32029) is the only dust-producing star with a metallicity below [Fe/H] = --1.45, whereas only 21\% of cluster stars have [Fe/H] $>$ --1.45 \citep{JP10}. It may be that the metal-poor population possesses a sufficiently high gas-to-dust ratio that dust is not important in their winds. We note that the aforementioned temperature selection bias also biases us towards metal-rich stars, which are cooler at a given luminosity.

(3) The dominant form of dust produced in the cluster is likely to be metallic iron, followed by silicates. Only the two most metal-rich stars have visible silicate emission features, compared to five other (still comparatively metal-rich) stars that produce solely metallic iron. The availability of iron to produce such opacity is discussed in Section \ref{ImplySect}.

As we will discuss in the next section, the inaccuracies in determining mass-loss rates with absolute accuracy means that we can neither determine the true composition of the dust produced within the cluster, nor the cluster's dust-production rate. We note, however, that the mass-loss rates for individual stars in Table \ref{DustyTable} have increased by a factor of $\sim$6 from those listed in MvLD. The summed dust production rate of these stars is approximately 2$\times$ that listed in MvLD, principly due to the determination of V42 as metal-rich. We remind the reader that we have not necessarily observed all the cluster's dusty stars in this work.

\subsection{Implications for low-metallicity environments}
\label{ImplySect}

\begin{figure}
\centerline{\includegraphics[height=0.47\textwidth,angle=-90]{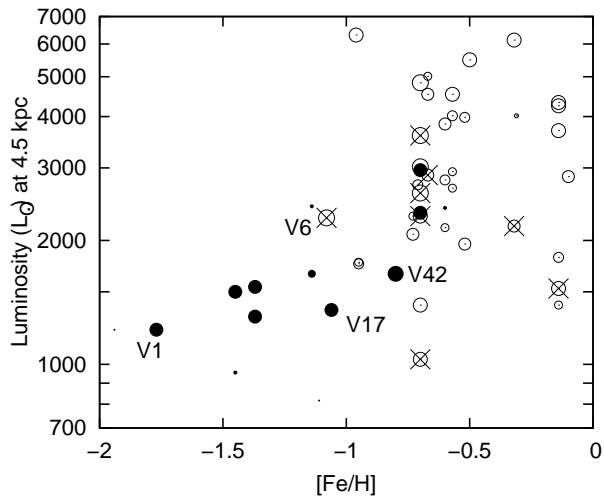}}
\caption{Variation of dust type among stars in 20 globular clusters. Additional data from \citet{SMM+10}; \citet{MSZ+10} and \citet{MBvLZ11} are shown. Stars with pure iron dust are shown as filled circles; those also forming silicates as open circles. Crosses show stars with crystalline features. Symbol sizes vary approximately with mass-loss rate.}
\label{MBolFeHFig}
\end{figure}

Fig.\ \ref{MBolFeHFig} shows $\omega$ Cen's stars in context with other globular cluster giants. Metal-poor ([Fe/H]$\lesssim$--1.2) stars appear to only produce iron dust, with silicates becoming more prevalent with increasing metallicity. Crystalline silicates appear confined to stars with lower luminosities and there is a lack of luminous metal-poor stars.

The low terminal velocities implied by our modelling are typically lower than the thermal speed of small molecules and the turbulent velocity in the wind, and much lower than the pulsation amplitude and escape velocity of the stars (Table \ref{DustyTable}). This problem has been well-noted in other globular clusters \citep{MvLD+09,BMvL+09,MSZ+10,MBvLZ11}. As gravity is not fully modelled in {\sc dusty}, it is not clear whether such outflows can be sustained. This implies that radiation pressure on dust is probably not the dominant method of accelerating dust from the star, therefore the wind velocity (and by implication, the mass-loss rate) may be higher than we model. Despite this, observations of low-metallicity stars \citep{MvLM+04} have so far followed the same velocity--luminosity--metallicity relation we use here (Eq. (\ref{VEqun}); see also \citealt{vanLoon00}). Velocities as low as 3 km s$^{-1}$ have been found in Halo carbon stars \citep{LZM+10}. However, it is not clear that effective dusty outflows can exist at lower outflow velocities, given the other physical mechanisms at work.

\citet{MBvLZ11} discuss possible means to modify the output wind parameters listed in Table \ref{DustyTable} by changing the input wind parameters (grain size, density and porosity; wind-driving mechanism; etc.), while keeping the mass-loss rate below limits implied by the pace of stellar evolution. In that work, the cluster 47 Tuc was examined. At [Fe/H] = --0.7, it is somewhat more metal-rich than $\omega$ Cen. Here, a combination of needle-like iron grains and a gas-to-dust ratio closer to unity were proposed to both increase wind velocity without substantially increasing mass-loss rate, and to bring the relative fractions of iron and silicates in the wind closer to canonical predictions for stars where silicates also condense. If one replaces the assumed spherical metallic iron grains with elongated cylinders of the same volume, one can more-efficiently accelerate the wind. Such grains would need to be substantially elongated: a typical axial ratio of $\gtrsim$12:1 is required to raise the velocity of V184's wind to above 3 km s$^{-1}$.

One can instead change the gas-to-dust ratio, leaving the grain shape spherical. As $v_\infty \propto \psi^{-1/2}$ (Eq.\ \ref{VEqun}), V184's gas-to-dust ratio would have to be $\gtrsim$38 times higher (i.e.\ $\psi = 125$) to attain a 3 km s$^{-1}$ outflow. Given complete condensation of all metals is attained for $\psi = 1230$, this would appear impossible. We therefore consider elongated grains to be more a more likely solution.

An alternative approach is to assume the iron and silicon lost from the star are entirely condensed into metallic iron and silicate grains. In this case, we would expect an iron:silicate ratio of approximately 1:2. If we assume identical, spherical grains (as above), we estimate a ratio of 4:1 for V6 (we ignore the small amounts of FeO here). If the spherical iron grains (but not the silicate grains) are replaced with needles, then the axial ratio must be $\gtrsim$16:1 in order to reproduce the expected ratio of 1:2. This approach, however, neglects important points such as differences in temperature between the grain populations (\S\ref{V6Sect}).

If the infrared excess is caused by AmC, rather than metallic iron, then the situation improves somewhat. Adopting AmC, mass-loss rates typically drop by a factor of 3, with wind velocities increasing by a similar factor. This would provide a method of attaining substantial wind velocities without unphysically-high mass-loss rates. There are substantial difficulties in producing AmC in an oxygen-rich environment, however, as all available carbon is thought to be rapidly sequestered into CO (e.g.\ \citealt{FG02}). Recent spectroscopic abundance determinations have also shown carbon to be even more depleted than iron in $\omega$ Cen's stars, including the slightly-dusty V394 and the `naked' LEID 61015 \citep{SDCN10}. This further limits the ability of the wind to produce AmC and suggests that the iron abundance, rather than carbon abundance, controls the production of dust in these stars. The iron abundance will hence set the fraction of dust in the wind, $\dot{D} / \dot{M} = 1/\psi$, and the final outflow velocity, $v_\infty$. We show in Section \ref{BiSect}, on the other hand, that the fundamental mass-loss rate, $\dot{M}$ is set by conditions nearer the stellar surface.

\section{Conclusions}
\label{ConcSect}
We present the first large-scale mid-infrared spectral survey of a metal-poor ([Fe/H] $<$ --1) globular cluster ($\omega$ Centauri) and compare these infrared observations to high-resolution H$\alpha$ spectra.

Dust emission is detected around seven of a diverse sample of 14 stars. Dust production seems confined to the cluster's 20\% coolest, most-metal-rich stars and those (namely V1) that have evolved off the AGB. V42 (LEID 44262), a very strong mass loser, may belong to the anomalous, metal-rich ([Fe/H] = --0.8) population.

Assuming the featureless mid-infrared excess seen in these stars is due to circumstellar metallic iron dust grains, metallic iron is the dominant dust product in our sample and by implication in other low-metallicity stars. Only the metal-rich stars V6 (LEID 33062) and V17 (LEID 52030) produce silicates. V6 also shows evidence for crystalline silicate production. The low terminal velocities modelled for these winds make it seem unlikely that radiation pressure on dust grains is driving the stellar winds in these cases.

H$\alpha$ observations of 12 of the 14 observed stars show significant outflow in nine targets. The remaining three stars (V6, V17 and V42) are very strong pulsators. Shock excitation of circumstellar gas caused by this pulsation precludes accurate measurements of their chromospheric outflows. This shows that substantial mass loss is ubiquitous among stars near the RGB/AGB tips, though dust production is not. While dust may be the most visible tracer of late-stage stellar mass loss, the onset of its production may not trace a fundamental change in the star itself, but is instead a byproduct of conditions set in the wind.


\section*{Acknowledgements}

RDG was supported by NASA and the United States Air Force. CEW was supported in part by the National Science Foundation, and NASA JPL/Spitzer grants awarded to the University of Minnesota. CIJ was supported by the National Science Foundation under award No. AST-1003201.



\label{lastpage}

\end{document}